\documentclass[aps,prb,reprint,twocolumn,showpacs,superscriptaddress]{revtex4-2}
\usepackage{amsmath}
\usepackage{graphicx}
\usepackage{lmodern}
\usepackage{amsmath}
\usepackage{color}
\usepackage{amssymb}
\usepackage{bm}
\usepackage{braket}
\usepackage{mathtools}
\usepackage{afterpage}
\usepackage{soul}
\usepackage[caption=false]{subfig}
\newcommand{\bea}{\begin{eqnarray}}
\newcommand{\eea}{\end{eqnarray}}

\newcommand{\expval}[1]{\langle{#1}\rangle}

\renewcommand{\(}{\left(}
\renewcommand{\)}{\right)}

\renewcommand{\]}{\right]}

\newcommand{\be}{\begin{equation}}
\newcommand{\ee}{\end{equation}}
\newcommand{\bpm}{\begin{pmatrix}}
\newcommand{\epm}{\end{pmatrix}}

\usepackage[dvipsnames]{xcolor}
\usepackage[colorlinks=true]{hyperref}
\hypersetup{
    colorlinks=true,
    linkcolor=blue,
    citecolor=blue,
    urlcolor=blue
} 

\makeatletter
\def\maketitle{
\@author@finish
\title@column\titleblock@produce
\suppressfloats[t]}
\makeatother

\begin{document}
\title{\texorpdfstring{Many-body higher-order topological invariant for $C_n$-symmetric insulators}{}}
\author{Ammar Jahin}
\affiliation{Department of Physics, University of Florida, Gainesville, Florida 32611, USA}
\author{Yuan-Ming Lu}
\affiliation{Department of Physics, The Ohio State University, Columbus, 43210, USA}
\author{Yuxuan Wang}
\email{yuxuan.wang@ufl.edu}
\affiliation{Department of Physics, University of Florida, Gainesville, Florida 32611, USA}

\date{\today}

\begin{abstract}
Higher-order topological insulators in two spatial dimensions display fractional corner charges. While fractional charges in one dimension are known to be captured by a many-body bulk invariant, computed by the Resta formula, a many-body bulk invariant for higher-order topology and the corresponding fractional corner charges remains elusive despite several attempts. Inspired by recent work by Tada and Oshikawa, we propose a well-defined many-body bulk invariant for $C_n$ symmetric higher-order topological insulators, which is valid for both non-interacting and interacting systems. Instead of relating them to the bulk quadrupole moment as was previously done, we show that in the presence of $C_n$ rotational symmetry, this bulk invariant can be directly identified with quantized fractional corner charges. In particular, we prove that the corner charge is quantized as $e/n$ with $C_n$ symmetry, leading to a $\mathbb{Z}_n$ classification for higher-order topological insulators in two dimensions.
\end{abstract}
\maketitle

\section{Introduction}
Protected by crystalline symmetries, higher-order topological insulators (HOTI) exhibit nontrivial gapless or degenerate states at the higher-order boundaries of a sample, such as corners in two spatial dimensions and hinges in three spatial dimensions~\cite{Hughes_2017,Benalcazar_2017,Schindlereaat_2018,Shapourian_2018,Piet_2017,Piet_2019,Eslam_2018,Das_Sarma_2019b,Katsunori_2019,Zhongbo_2019b,Dai_Xi_2019,Vladimir_2019,Qiang_Wang_2018,Fulga_2018,Sun_2020,Penghao_2020,Piet_2018,Ezawa_2018,Fan_2019}. 
As a paradigmatic example, the Benalcazar-Bernevig-Hughes (BBH) model~\cite{Hughes_2017,Benalcazar_2017}, defined on a square lattice, is fully gapped in the bulk and at the boundary, but exhibits fractional charges $\pm e/2$ at the corners. 
More generally, in the presence of $C_2$, $C_3$, and $C_6$ symmetries, by directly constructing lattice models, it has been shown that  HOTI's exhibit richer patterns of quantized fractional corner charges~\cite{Benalcazar_2019}. 
Beyond HOTIs, the concept of higher-order topology has since been extended to superconductors~\cite{Yuxuan_2018,Sayed_Ali_2018,Bohm_Jung_2019,Zhongbo_2019,Das_Sarma_2019a,Zaletel_2019,Das_Sarma_2019c,Das_Sarma_2020,bitan_2020,Wang_Zhong_2018,Wieder_2020}, semimetals~\cite{Mao_2018,Duan_2023,Jiang_2023,Chen_2023,Zhou_2023,Ming_2023}, and intrinsically interacting bosonic systems in two and three spatial dimensions~\cite{Serra_Garcia_2018,Peterson_2018,Mittal_2019,El_Hassan_2019,Chen_2020,He_2020,Zhengyou_2023,Medina_2023}.

Via the bulk-boundary correspondence, the nontrivial boundary states for HOTI's should be naturally captured by topological invariants in the bulk. 
In the BBH model, multiple topological invariants have been proposed, such as the nested Wilson loop in the presence of mirror symmetries~\cite{Hughes_2017,Benalcazar_2017} and eigenvalues of the rotation operator at high symmetry points in the Brillouin zone. 
However, these invariants are based on band structures and there is no straightforward generalization to interacting systems. 
Furthermore, as pointed out in Refs~\cite{Benalcazar_2017,Ono_2019}, the nested Wilson loop is not protected by a bulk gap closing, thus making it difficult to interpret as a bulk topological invariant.
A $\mathbb{Z}_4$ Berry phase topological invariant was proposed in Ref.~\cite{similar_inv_2020} for the BBH model, but a bulk-boundary correspondence was not generally established, and it is not clear how to generalize to different models.

For topological insulators in one spatial dimension (1d) hosting $\pm e/2$ fractional charges at the ends,  a bulk many-body topological invariant applicable in the presence of interactions has been long known as the Resta formula, which captures the quantized electric polarization density in the bulk. By analogy, it has been proposed~\cite{Wheeler_2019,Kang_2019} that $C_4$ symmetric 2d HOTI's are characterized by a quantized bulk quadrupole density, related to the phase of the expectation value of the operator
\be
\hat U_{xy}= \exp \(\frac{2\pi i}{L^2} \sum_j \hat n_j x_j y_j\),
\ee
where $L$ is the system size, $\hat n_j$ is the electron density operator at site $j$.

However, such a straightforward generalization of the Resta formula has been shown to be problematic~\cite{Ono_2019}, as it violates the periodic boundary condition and does not always lead to quantized values. The authors of Ref.~\cite{Wheeler_2019} showed that this pathology is related to the nonzero dipolar \emph{flucutations} of the system, and is resolved if one instead considers a limit in which the system has an additional dipole conservation symmetry. Unfortunately, for a generic system, especially in the presence of interactions, there is no generic procedure to take such a dipolar symmetric limit without closing the bulk gap. Finally, it is unclear how a bulk quadrupole density can account for various quantization patterns for $C_n$ symmetric HOTI's, particularly since the crystalline symmetries do not play any role in the construction of the proposed many-body invariant.

In a recent work, Tada and Oshikawa~\cite{Tada_2023} proposed a new many-body invariant for $C_4$ symmetric HOTI's. The authors noted that when the system is subject to a $2\pi$ flux, under the Landau gauge, while $\hat C_4$ is no longer a symmetry, $\hat {\tilde C}_4 \equiv \hat C_4 \hat U_{xy}$ is. It was then suggested that $e^{2\pi i q}\equiv \langle \Psi_{2\pi} |\hat {\tilde C}_4 |\Psi_{2\pi}  \rangle /\langle  \Psi_{0}| \hat {\tilde C}_4 | \Psi_{0}\rangle$, in which $|\Psi_{2\pi,0}\rangle$ are the ground state wave functions with or without a $2\pi$ flux, can serve as a many-body topological invariant of the system. The authors argued that this invariant is related to the bulk quadrupole moment captured by $\hat U_{xy}$, but is a well-defined quantum number and free from the aforementioned issues. The authors further argued that a nontrivial value taken by this invariant indicates ground state degeneracy caused by the corner states. However, the direct relation between the quantized fractional corner charge and the bulk invariant was not established. 

A natural question is whether a similar many-body invariant can be defined by generalizing this idea, and whether there exists a direct correspondence between the bulk invariant and the fractional corner charge. As we mentioned above, this may require going beyond the paradigm of bulk multipolar density. 
For example, it has been shown that for $C_6$ invariant HOTI, the fractional part of corner charges can take values of multiples of $\pm e/6$~\cite{Benalcazar_2019}. 
Depending on the arrangement of the charges on the six corners, it is not clear how a single bulk multipolar moment can adequately account for the corner charge patterns.
In fact, we show in Sec.~\ref{sec:no_quad_moment} that while the corner charges can lead to a planar electric octupole moment for a macroscopic hexagonal sample, this total moment cannot come from integrating over a bulk octupolar density.

In this paper, we address these two issues for a $C_n$-symmetric HOTI. Modifying and extending the results in Ref.~\cite{Tada_2023}, we provide a many-body bulk topological invariant $q$ satisfying $nq\in \mathbb{Z}_n$  defined 
  under periodic boundary conditions (topologically equivalent to a torus). Defined in Eq.~\eqref{eq:1}, the topological invariant $q$ is obtained via the ratio of $C_n$ eigenvalues of the ground state with or without a background $2\pi$ magnetic flux through the macroscopic sample. As the associated magnetic field is infinitesimal, this invariant probes the property of the ground state, protected by a finite energy gap. For $n=4$, in the Landau gauge, our invariant reduces to that proposed in Ref.~\cite{Tada_2023}.

Compared with Ref.~\cite{Tada_2023}, our result and its interpretation do not rely on the connection with a bulk multipolar moment. In fact, we argue that the resemblance of the many-body invariant proposed in Ref.~\cite{Tada_2023} with that in Refs.~\cite{Wheeler_2019,Kang_2019} based on electric quadrupole moment is an artifact of choosing the Landau gauge. Instead, our work establishes directly a relation between $q$ and the fractional part of the corner charge $\nu$ for a $C_n$-symmetric HOTI in 2d. To that end, the physical meaning of the bulk invariant becomes clear as we adiabatically turn the hopping amplitudes $t'$ across boundaries of the system to be parametrically weaker than the bulk, which is $\sim t$. As long as $t'$ remains nonzero, the bulk gap remains open and the quantized value $q$ cannot change. We show that in the spirit of the renormalization group (RG), the effective low-energy Hamiltonian is captured by the hopping and interaction among localized states near the corners of the system. By matching the low-energy (IR) and high-energy (UV) data of the theory, we show that $q$ effectively represents the fractional part of the electron filling per corner. As one sets $t'=0$, the boundaries become open, and the system host degenerate localized corner states. Under open boundary conditions, we obtain the fractional part of the electric charge $\nu$ at each corner, which also accounts for the ionic contribution, and relates it to $q$ via $\nu = q \mod 1$ (in units where the electron charge $e=1$).

Importantly, the many-body invariant and the bulk-boundary correspondence remain well-defined in the presence of interaction effects. Thus, we conclude that $C_n$-symmetric second-order topological insulators are classified by $\mathbb{Z}_n$. For crystalline systems, $n$ can only take values of $2,3,4,6$, but for noncrystalline systems (e.g., quasicrystals), $n$ can take all integer values. Unlike previous attempts by generalizing bulk multipolar moments,  our construction does not rely on translation symmetry, and thus we expect our many-body invariant to be applicable to quasicrystals with a generic $n$. Interestingly, corner charges of $e/4$ have been obtained in a twisted bilayer graphene structure with a twist angle of $\theta = \pi/6$~\cite{Nigel_2020}. Such a system has a $C_{12}$ symmetry, and our topological invariant with $n=12$ is indeed compatible with an $e/4$ corner charge.

{Recently, classification of $C_n$ symmetric topological phases in 2d has been studied by several authors~\cite{Peng_2019, khalaf2019, yizhi2020, Manjunath2021, maissam1, maissam2, maissam3, maissam4, Mayman_2022,  jonah2022, Manjunath2023} from the perspective of generalized Wen-Zee effects~\cite{WenZee1992}, which describe the mixed anomaly between charge symmetry and spatial rotational symmetry. We show that our bulk invariant is proportional to the generalized Wen-Zee shift $\mathcal{S}$, and thus the  classification results from the two approaches are consistent. Compared with these field-theoretic or topological quantum chemistry~\cite{jonah2022} approaches, our proof from UV-IR mapping directly relating the bulk invariant to the second-order topology and corner charge is novel.}


We demonstrate the results for our many-body invariant for several lattice models on square and honeycomb lattices. These lattice models have been shown~\cite{Benalcazar_2019} to exhibit various fractional corner chargers within some certain parameter range. Indeed, the corner charge agrees with a direct calculation of the many body invariant, which can only change upon a gap closing in the bulk. We expect the agreement to hold quantitatively even in the presence of interaction effects.

The rest of this paper is organized as follows. In Sec.~\ref{sec:no_quad_moment} we argue that in general the corner charges in a $C_n$ symmetric higher-order topological insulator cannot be described by a bulk electric multipolar moment. In Sec.~\ref{sec:def} we define our many-body topological invariant via the change of $C_n$ eigenvalue upon adding a flux quantum. In Secs.~\ref{sec:proof1} and \ref{sec:proof2} we provide two different proofs directly relating our topological invariant to the fractional corner charge of the system under open boundary condition. In Sec.~\ref{sec:ex} we show several examples in which we numerically compute the topological invariant as well as their corresponding fractional corner charges, which remain consistent across topological phase transitions.

\section{Inadequacy of a bulk quadrupole moment} \label{sec:no_quad_moment}

\begin{figure}
    \centering 
    \includegraphics[scale=0.8]{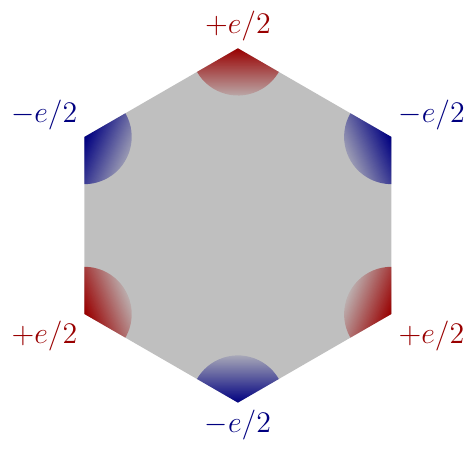}
    \caption{The charge distribution of a higher-order topological insulator protected by $C_6$ with alternating $\pm e/2$ charges on a hexagonal sample. Trying to understand the charge distribution as a direct consequence of an octopole moment in the bulk is not possible as the corner charges do not scale with the system size.}
    \label{fig:hexa_pm_charges}
\end{figure}
Before we discuss the many-body invariant, let us begin by explaining why a bulk multipolar density is inadequate to account for the corner charges in a $C_6$-symmetric HOTI.

As was shown in Ref.~\cite{Benalcazar_2019} by a concrete lattice model (see Section \ref{sec:ex} for more details), a $C_6$-symmetric HOTI can host $\pm e/2$ charges at the six corners.
When the charges alternates in sign every $\pi/6$ (see Fig.~\ref{fig:hexa_pm_charges}), the system displays a macroscopic electric octupole moment. According to basic electrodynamics, the octupolar moment is expressed as
\be
\mathcal{O} = \int d\bm r\, \rho(\bm {r})(3x^2-y^2)y = - \frac{e}{2} \times 6L^3,
\ee
where $\rho(\bm r)$ is the electric charge density, and $L$ is the length of the side of the hexagon.

Note that $\mathcal{O} \sim L^3$ for the 2d system in consideration. This immediately indicates that a bulk octupole density, which is an intensive quantity, cannot account for the macroscopic octupolar moment. Indeed, a bulk density can only lead to an octupolar moment $\sim L^2$, and thus a vanishingly small corner charge in the thermodynamic limit.

Therefore, we conclude that in general quantized corner charges of a  $C_n$-symmetric HOTI do not correspond to the multipolar density in the bulk.

\section{Definition of the many-body bulk invariant}\label{sec:def}

The system is defined on a macroscopic $\tilde n$-sided regular polygon ($\tilde n$-gon) with opposite sides identified. 
With $C_n$ symmetry, $\tilde n\geq 4$ is chosen to be a multiple of $n$. For example, for $n=3$, we can choose $\tilde n = 6$.
As can be checked from its Gauss character, this is topologically equivalent to a torus for all $m$. Under periodic boundary conditions, due to the Dirac quantization condition, the magnetic flux through the torus can only be integers of $2\pi$, which we choose to be either $0$ and $2\pi$ here.

\begin{figure}
  \includegraphics{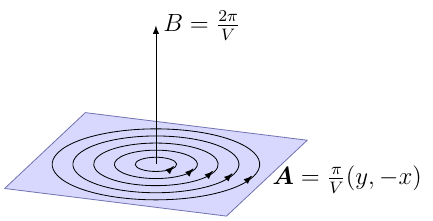}
  \caption{Adding a magnetic field perpendicular to a $C_n$ symmetric system using the symmetric gauge does not spoil the $C_n$ symmetry. }
  \label{fig:sym_gauge}
\end{figure}

Under symmetric gauge $\bm A \propto (y, -x)$ in the bulk, as shown in Fig.~\ref{fig:sym_gauge}, the system remains $C_n$ symmetric with or without the $2\pi$ flux.
The topological invariant is then defined as
\be
\exp({2\pi i q}) = \frac{\langle \Psi_{2\pi}|C_n|\Psi_{2\pi} \rangle}{\langle \Psi_{0}|C_n|\Psi_{0} \rangle},
\label{eq:1}
\ee
where $\Psi_{0,2\pi}$ are the ground states of the Hamiltonian in the absence/presence of a uniform $2\pi$ flux through the torus. In the thermodynamic limit, this represents a vanishingly weak magnetic field, and the ground states belong to the same phase protected by the insulating gap.
The many-body Hamiltonian in the presence of the $2\pi$ flux is obtained via Peierls substitution to the background gauge field $\bm A$. Since the $2\pi$ flux can be viewed as coming from a magnetic monopole inside the torus, the $\bm A$ field cannot be covered by a single gauge choice. In particular, the hopping terms across the boundary of the $\tilde n$-gon needs to be determined separately. 

\begin{figure}
  \centering
  \subfloat[]{\includegraphics{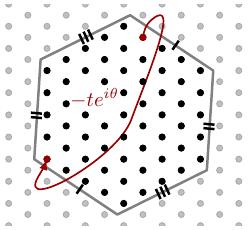}}
  \hfill
  \subfloat[]{\includegraphics{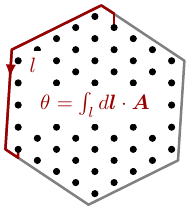}}
  \caption{We calculate the invariant on a finite system defined on a torus. In the $C_6$ case, we obtain the torus by identifying opposite sides of the hexagon defining the boundaries of the system as shown in (a). This identification means that hopping is allowed between unit cells on opposite boundaries. An example of such hopping is shown in red in (a).  In the presence of the vector potential, hoppings crossing the boundary obtain a phase $\theta$ that is the line integral of the vector potential on the contour shown in (b). For other values of $n$, the values of hopping amplitudes across boundaries can be obtained similarly.}
  \label{fig:def_of_oboundary_phase_hop}
\end{figure}

\begin{figure}
  \centering 
  {\includegraphics[scale=1]{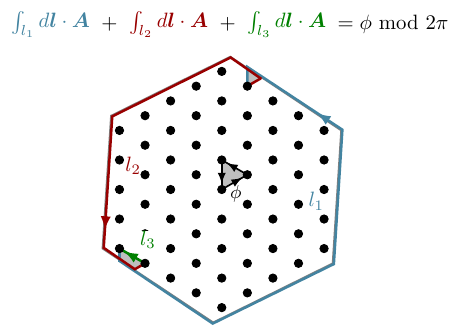}}
  \caption{The modified boundary hoppings are designed to ensure the flux inside the boundary plaquettes is the same as the bulk plaquettes $\phi$. }
  \label{fig:bounday_plaq_flux}
\end{figure}

Here we introduce a consistent way to directly determine the phases of hopping amplitudes across the boundaries that gives the correct value for flux through loops across a boundary and preserves the $C_n$ rotation symmetry. As shown in Fig.~\ref{fig:def_of_oboundary_phase_hop} (with $n=\tilde n=6$), the destination of hopping across the boundary is determined by the periodic boundary condition of the $\tilde n$-gon (the torus). The phase of the complex hopping amplitude is determined by the line integral of the gauge potential $\bm A$ on a path that involves a counter-clockwise circulation around the boundary. By construction, the system remains $C_n$ symmetric, and as can be seen in Fig.~\ref{fig:bounday_plaq_flux}, any loop across the boundary encloses the same flux as it would in the bulk, modulo $2\pi$.

\label{sec:def_inv}

Since $(C_n)^n$ is proportional to the identity operator, the eigenvalues of $C_n$ are quantized as $\lambda_m = \exp(2\pi i m/n) \lambda_0$. As a result,  we conclude that $C_n$-symmetric HOTI's in 2d are classified by 
\be
nq\in \mathbb{Z}_n.
\label{eq:qzn}
\ee
We note that the $\mathbb{Z}_4$ classification for $C_4$ symmetric HOTI in 2d has been shown using a Berry phase topological invariant in Ref.~\cite{similar_inv_2020} specifically for the BBH model.

\section{Bulk-boundary correspondence from UV-IR mapping}\label{sec:proof1}
In this Section, we prove that the topological invariant in Eq.~\eqref{eq:1} directly corresponds to the fractional charge localized at the corners of a $C_n$-symmetric 2d insulator. {To be precise, throughout  this work, by ``corners", we refer to a set of $n$ elements related by $C_n$ symmetry that form its regular representation. This includes cases with $\tilde n \neq n$, where the combination of $\tilde n/n$ neighboring vertices of the $\tilde n$-gon is viewed as a proper corner. Indeed these vertices can geometrically merge into one single corner without breaking $C_n$ symmetry.}

\begin{figure}
    \centering 
    \includegraphics[scale=0.95]{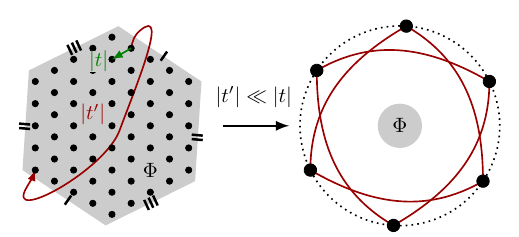}
    \caption{Let the magnitudes of hoppings crossing boundaries be $|t'|$ and those in the bulk be $|t|$ as shown in the figure on the left. We can take open boundary conditions by setting $|t'| = 0$. In this limit, we expect the system to host corner charges. Studying the system with $|t'| \ll |t|$, the model representing the low energy physics is that of particles hopping on a ring as shown in the figure on the right. If the original system has a flux $\Phi = 0, 2\pi$ piercing the bulk, the low energy ring also has the same flux going through it. 
    The ground state of the low energy model after threading the flux is related to the ground state before threading the flux by a large gauge transformation.}
    \label{fig:4}
  \end{figure}

The corner excitations can be made energetically distinct from the bulk by setting all hopping amplitudes across edges and corners $t'$ to be much smaller than their bulk counterpart $t$,~\footnote{To display corner fractional charge in open boundary condition, in some cases one would need to add an onsite potential to the corner and/or the edge to ensure the chemical potential passes the corner modes. In those cases, for our proof we would also need to add the same potential in addition to taking $t'\ll t$.}  as shown in Fig.~\ref{fig:4}. In the process of decreasing $t'$ the energy gap is kept open, and thus the value of $e^{2\pi i q}$ cannot change, whose $n$-th power must be one.  In an RG sense, as the energy scale of the system flows from $t$ to $t'\ll t$, the system can be described by an effective IR Hamiltonian. In the absence of gapless edge states when $t'=0$, we expect the Hilbert space of the IR Hamiltonian to be supported on the corners. The IR Hamiltonian describes hopping among $n$-sites forming a ring (see Fig.~\ref{fig:4}; the exact hopping pattern depends on $n$ and is not important for our purposes). The $C_n$ symmetry for the entire system gets mapped to discrete translation symmetry $T_n$ of the ring.

Now, the bulk invariant \eqref{eq:1} computed using the UV Hamiltonian can be expressed purely by physics in the IR, which ultimately describes the same ground state. The IR physics is much simpler --- in particular, adding a $2\pi$ flux to the bulk is simply captured by a large gauge transformation to the $n$-site ring. Matching UV with IR, we have the following identification:
\begin{align}
C_n\sim T_n e^{2\pi i L[\bm A]/n}, |\Psi_{0}\rangle \sim |\psi_0\rangle,  |\Psi_{2\pi}\rangle\sim U|\psi_0\rangle,
\label{eq:4}
\end{align}
where $T_n$ is the translation operator of the $n$-site boundary system, $L[\bm A]$ is an integer c-number corresponding to the total angular momentum from UV (bulk) fermionic modes, $|\psi_0\rangle$ is the ground state of the IR Hamiltonian with and without the $2\pi$ flux, and $U$ is the large gauge transform operator given by 
\bea
\hat U = e^{\frac{2\pi  i}{n}\sum_{j=1}^n \hat n_j j},
\eea
where $\hat n_j$ is the electron number operator at corner $j$, $1\leq j<n$, and we have set $e=1$.

First, we show that in the thermodynamic limit, The angular momentum from the bulk $L[\bm A]$ is independent of $\bm A$, i.e., does not change upon on the insertion of the $2\pi$ flux. We express the bulk operator $\hat L[\bm A]$ as 
\be
\hat L[\bm A] = \frac{1}{n}\sum_{\bm r \in \mathrm{bulk}} \sum_{\ell = 0}^{n-1} \ell\, c^\dagger_\ell(\bm r) c_{\ell}(\bm r),
\ee
where $c_{\ell}(\bm r)$ is an annihilation operator that transforms as a 1d irreducible representation of the $C_n$ group:
\be
c_{\ell} (\bm r) = \frac{1}{\sqrt{n}}\sum_{j=1}^{n} \exp{\frac{2\pi i\ell j}{n}} c((\mathsf C_n)^j \bm r)
\ee
where $c(\bm r)$ is the annihilation operator at $\bm r$ comprising only bulk modes, and  $(\mathsf C_n)^j \bm r$ is the coordinate of a site at $\bm r$ subject to $j$ consecutive $C_n$ rotations. The expectation value $L[\bm A]$ can be obtained using Green's function. We have
\begin{align}
L[\bm A] =& \langle \hat L[\bm A] \rangle = \sum_{\bm r\in \mathrm{bulk}}  \sum_{\ell=0}^{n-1}\sum_{j=1}^{n} \frac{\ell}{n} e^{2\pi i \ell j/n} G_{\bm A} (\bm r, (\mathsf C_n)^j \bm r), 
\label{eq:8}
\end{align}
where 
\be
G_{\bm A} (\bm r, \bm r') \equiv \langle \Psi_{\bm A}|c^\dagger(\bm r)  c(\bm r')| \Psi_{\bm A} \rangle.
\ee
For any bulk sites $\bm r$, it is straightforward to see that $G_{\bm A}(\bm r, \bm r) =n(\bm r)$, i.e., the electron density at site $\bm r$, which is independent on the magnetic flux and $\bm A$. For $j\neq 0$ we have for a bulk insulator
\be
G_{\bm A}(\bm r, (\mathsf C_n)^j \bm r) \propto \delta_\xi(\bm r - (\mathsf C_n)^j \bm r) \exp\({i\Phi_{\bm A} \frac{r^2}{nL^2}}\).
\ee
Here $L$ is the system size, $\Phi_{\bm A}$ is the magnetic flux through the system, and $\delta_\xi(\bm r)$ is a wave packet rapidly decaying beyond a characteristic length $\xi$. The decaying behavior of $G_{\bm A}$ follows from the fact that the system is an insulator in the bulk with a correlation length $\xi \ll L$. Importantly, since the fermions are projected to only include bulk modes, the correlation function has no ``shortcuts" across the corners, and thus $G_{\bm A}(\bm r, (\mathsf C_n)^j \bm r)$ is significant only for $\bm r$ near the origin. The presence of a magnetic flux  $\Phi_{\bm A}$ introduces an additional phase via (the shortest) path connecting $\bm r$ and $(\mathsf C_n)^j \bm r$. 
 However, in the thermodynamic limit $L\gg \xi$, $\delta_\xi(\bm r)$ can be replaced by $\delta(\bm r)$ for the summation in Eq.~\eqref{eq:8}. We have 
 \be
 L_{\bm A} \propto \frac{N(n-1)}{2},
 \ee
where $N$ is the total number of electrons. We see that the $\bm A$ dependence vanishes. For this reason, using the identification in Eq.~\eqref{eq:4}, we can rewrite the topological invariant $e^{2\pi i q}$ in terms of IR quantities as
\be
\exp({2\pi i q}) = \frac{\expval{\psi_{0}|U^\dag T_nU|\psi_{0}}}{\expval{\psi_0|T_n|\psi_0}}.
\ee
Effectively, we have deformed the system in consideration from a 2d sample to a ring with $n$ sites with or without a $2\pi$ flux penetrating it, shown in Fig.~\ref{fig:4}. As we explained, the main advantage is that the wave functions of the ground state with or without the $2\pi$ flux simply differ by a large gauge transformation $U$.

The operators for the $n$-site ring have the following algebra
\be
T_n \hat n_j T_n^{-1}= \hat n_{j+1},
\ee
which leads to
\bea
U^\dagger T_nU T_n^{-1}=e^{-2\pi i \sum_j\hat n_j/n}\equiv e^{-2\pi i\hat\rho},
\eea 
where $\hat \rho\equiv \sum_j \hat n_j /n$ is the average electron number operator for each corner. Thus the topological invariant evaluates to 
\bea\label{2pi_flux_Cn_1}
\exp({2\pi i q}) =\frac{\expval{\psi_{0}|U T_nU^\dagger|\psi_{0}}}{\expval{\psi_0|T_n|\psi_0}}=\langle \psi_0| e^{-2\pi i\hat\rho}|\psi_0\rangle.
\eea
Notice that since there are nonvanishing hopping amplitudes between the corners ($t'$), the ground state $|\psi_0\rangle$ is not an eigenstate of the particle number $\hat n_j$ on each site. However, since UV and IR degrees of freedom decouple in the $t'\ll t$ limit, the state $|\psi_0\rangle$ involving only IR degrees of freedom (the corner electrons) should conserve particle number. Therefore $\sum_{j}\hat n_j$ is a good quantum number and so is $\hat \rho$. We can thus equate the right hand side of Eq.~\eqref{2pi_flux_Cn_1} with $e^{2\pi i \rho}$, where $\rho = \langle \hat \rho \rangle \mod 1$ is the fractional part of the average electron number at the corners.

We see that our topological invariant defined in Eq.~\eqref{eq:1} satisfies
\bea\label{eq:18}
  q= -\rho  \mod 1
\eea
i.e., captures the fractional filling of the ring formed by $n$ corner states. Incidentally, we note that this derivation using the braiding algebra between translation and large gauge transformation is quite similar to the non-perturbative derivation of the Luttinger's theorem in 1d~\cite{oshikawa-lutt, ersatz}, which formally corresponds to the $n\to \infty$ limit.

The corner charge for a HOTI is computed under open boundary condition, i.e., $t'=0$. To this end, we note that the ground state $|\psi_0\rangle$ can be written as a superposition of localized states $|\varphi_i\rangle$ without entanglement among the corners, all of which have the same total corner electron number $n\rho$.  When $t'=0$, all $|\varphi_i\rangle$ states become degenerate ground states. However, for all $|\varphi_i\rangle$, the electron occupation numbers for each corner can only take integer values, e.g., either 0 or 1. In a HOTI, the corner charge comes from the combination of both electrons and ions when $t'=0$. We know that for a tight-binding model with periodic boundary conditions and $t'\neq 0$, each corner is electrically neutral, even if the electrons contribute a fractional charge $\rho$ to each corner. After all, the electronic charge is compensated by ions in a tight-binding model.  This indicates that when corner electron filling becomes integer-valued under open boundary condition, there must be a fractional net charge $\nu$, which amounts to the additional electronic contribution between integer and fractional filling. We have
\be
\nu = -\rho  = q \mod 1,
\label{eq:17}
\ee
where the charge of an electron is set to $e=1$. This completes our proof that the bulk topological invariant in Eq.~\eqref{eq:1} directly captures the fractional part of the corner charge under open boundary condition. Just like $q$, the fractional charge $\nu$ is quantized by $1/n$ of the electron charge.

We emphasize that our proof applies to both interacting and non-interacting cases. Indeed, our proof only used the local nature of the single-particle Green's function in the bulk, and the algebraic relations between $T_n$ and $U$, none of which relies on the non-interacting limit. Finally, we note that our proof above does not rely on translation symmetry in the bulk either. This indicates that the $\mathbb{Z}_n$ invariant can be extended to non-crystalline systems with a generic value of $n$. As we mentioned in the Introduction, it has been obtained that a model defined in a twisted bilayer graphene structure with a twist angle $\theta=\pi/6$ and a $C_{12}$ rotation symmetry displays an $e/4$ corner charge. From Eqs.~(\ref{eq:qzn}, \ref{eq:17}), indeed this is compatible with our topological invariant. It would be interesting to directly compute it for the $C_{12}$-symmetric model in Ref.~\cite{Nigel_2020}, although it remains to be seen how such a noncrystalline system may be placed on a torus.

\section{Relation to the Wen-Zee term}\label{sec:proof2}

{Recently, several works~\cite{khalaf2019, Peng_2019, yizhi2020, Manjunath2021, maissam1, maissam2, maissam3, maissam4, Mayman_2022, jonah2022, Manjunath2023} have studied 2d $C_n$-symmetric topological phases using a topological responses which include the crystalline generalization of the Wen-Zee term. Here we briefly summarize the key results, and then demonstrate the complete agreement between our approach of UV/IR mapping for Eq.~\eqref{eq:1} and the approach from generalized Wen-Zee effect.}


The low-energy physics of a two-dimensional (2d) symmetry protected TCI is described by the Abelian Chern-Simons theory\cite{WenZee1992,Mayman_2022,Peng_2019,Levin2012,Lu2012}:
\begin{align}\notag
\mathcal{L}_\text{bulk}=&-\frac{\epsilon^{\mu\nu\lambda}}{4\pi}K_{I,J}a_\mu^I\partial_\nu a^J_\lambda\\
&+ \frac{\epsilon^{\mu\nu\lambda}}{2\pi}t_IA_\mu\partial_\nu a^I_\lambda+\frac{\epsilon^{i\mu\nu}}{2\pi}s_I\omega_i\partial_\mu a_\nu^I\label{eq:bulk action}
\end{align}
where $A_\mu$ is the external electromagnetic gauge field, and $\omega_i$ is the spin connection i.e. the gauge field associated spatial rotational symmetry $U(1)_r$ in 2d. We have followed the Einstein convention to sum over repeated indices. The $K$ matrix of a 2d TCI has the following general form\cite{Levin2012,Lu2012}
\bea
{K}=\hat 1_{N\times N}\otimes\bpm1&0\\0&-1\epm
\eea
where $\hat 1_{N\times N}$ is the identity matrix of $N$ dimensions. The $2N$-dimensional vector $\vec t=(t_1,\cdots,t_{2N})^T$ is known as the charge vector, while $\vec s=(s_1,\cdots,s_{2N})^T$ is known as the spin vector of a rotational-symmetric Abelian topological order~\cite{WenZee1992}. After integrating out the dynamical gauge fields $a^I_\mu$ which describes the low-energy topological excitations of the system, we achieve an effective field theory for quantized responses of the TCI:
\begin{align}
    \mathcal{L}_{\text{eff}} &= \frac{\sigma_{xy}}{4\pi}\epsilon^{\mu\nu\lambda} A_{\mu}\partial_{\nu} A_{\lambda} + \frac{\mathcal{S}}{2\pi}\epsilon^{\mu\nu\lambda} A_{\mu}\partial_{\nu} \omega_{\lambda} \nonumber \\&+ \frac{\ell_s}{4\pi}\epsilon^{\mu\nu\lambda} \omega_{\mu}\partial_{\nu} \omega_{\lambda},\label{Wen-Zee term}
\end{align}
where the Hall conductance $\sigma_{xy}$ (in unit of $e^2/h$) and the Wen-Zee shift $\mathcal{S}$~\cite{WenZee1992} are given by~\footnote{The $\ell_s$ term here does not take into account of the framing anomaly discussed in \cite{Gromov2015}.} 
\bea    
    \sigma_{xy} &= t^T K^{-1} t; \quad \mathcal{S} = s^T K^{-1} t; \quad \ell_s = s^T K^{-1} s.
\eea
In particular, the last term $\mathcal{L}_\text{Wen-Zee}=\frac{\mathcal{S}}{2\pi}A\text{d}\omega$ of the response theory (\ref{Wen-Zee term}), known as the Wen-Zee term~\cite{WenZee1992}, describes the quantized response from the mixed anomaly between global charge symmetry $U(1)_c$ and spatial rotational symmetry. {Initially proposed for systems with continuous rotation symmetry, the Wen-Zee term has been  extended to crystalline systems with discrete   $C_n$ rotation symmetries~\cite{Peng_2019, khalaf2019, Manjunath2021}.} {The key feature with  discrete $C_n$ rotation  is that the shift $\mathcal{S}$ is only well-defined modulo $n${, i.e., $\mathcal{S}\in \mathbb{Z}_n$, leading to the same classification as ours.}}

The Wen-Zee term can be interpreted in two ways as follows:

(i) In terms of orbital angular momentum (in unit of $\hbar$) carried by a fluxon excitation (i.e. a flux quantum) in the TCI phase~\cite{Manjunath2021}
\begin{align}\label{fluxon angular mtm}
\!\!\!\!\!L_z(\Phi)=\int \frac{\partial\mathcal{L}_\text{Wen-Zee}}{\partial\omega_0}=\int \mathcal{S}\frac{\partial_x A_y-\partial_y A_x}{2\pi}=\mathcal{S}\frac{\Phi}{2\pi}.
\end{align}
Specifically, a single fluxon ($\Phi=2\pi$) carries an orbital angular momentum of $L_z(2\pi)=\mathcal{S}$. As we mentioned, our topological invariant \eqref{eq:1} precisely measures this angular momentum, and thus 
\be\label{eq:25}
q= \frac{L_z(2\pi)}{n} = \frac{\mathcal{S}}{n}.
\ee

(ii) In terms of quantized charge (in unit of $e$) localized as a disinclination~\cite{Peng_2019,maissam1,Manjunath2023} 
\begin{align}\label{disclination charge}
\rho_{\rm dis}(\Theta)=\int \frac{\partial\mathcal{L}_\text{Wen-Zee}}{\partial A_0}=\int \mathcal{S}\frac{\epsilon^{0ij}\partial_i\omega_j}{2\pi}=-\mathcal{S}\frac{\Theta}{2\pi}.
\end{align}
since a disclination can be regarded as a curvature flux of the spin connection $\omega$that is equal to $-\Theta$, where $\Theta$ is the Frank angle.

Ref.~\cite{Manjunath2023} has recently shown that for an Abelian topological order with both charge conservation and $C_n$ rotational symmetry, when the system is placed on an $n$-gon, the corner charge $q$ is the same as a disclination with Frank angle $\Theta = 2\pi/n$.
This can be easily understood from the disclination charge formula (\ref{disclination charge}): for a finite sample to create a $\Theta=\frac{2\pi}{n}$ disclination on an $n$-gon, one has to remove a slice of the $n$-gon with a corner. {As the slice removed from the sample must have integer charge, the fractional charge hosted on the corner of the slice now must be trapped at the disclination. We then have}
\bea\label{eq:26}
\rho = \rho_{\rm dis} = -\frac{\mathcal{S}}{n}.
\eea
Combining Eqs.~(\ref{eq:25}, \ref{eq:26}), and noting that $\mathcal{S}$ is defined modulo $n$, we immediately obtain the bulk-boundary  correspondence \eqref{eq:18}.

\section{Examples}\label{sec:ex}
In this section, we numerically demonstrate the capability of our invariant to capture the topology of various $C_n$-symmetric insulators. For numerical simplicity, we focus on noninteracting cases, while we leave the numerical verification for interacting systems to future work.

For the fourfold symmetric case, we calculate the invariant for the BBH model at half and quarter-filling. 
The half-filling case has $e/2$ corner charges, and the quarter-filling case has $3e/4$ corner charges. 
For the sixfold symmetric case, we calculate the invariant for the two models that can generate all possible $C_6$-symmetric higher-order phases~\cite{Benalcazar_2019}. 
For a macroscopic hexagonal sample, these models host $e/2$ and $2e/3$ charges on the vertices. 
In the following, we introduce all the models and present the numerical evaluation of the proposed topological invariant in Eq.~\eqref{eq:1}.

All tight-binding models we study have the following structure
\begin{align}\label{eq:27}
    H(\bm A) = &-w \sum_{\bm R} \quad \  c^\dagger_{\bm R,\alpha} \  h_{\bm R \bm R}^{\alpha \beta}(\bm A)  \ c_{\bm R,\beta} \nonumber  \\ 
    &- t \sum_{\langle\bm R, \bm R'\rangle} c^\dagger_{\bm R,\alpha} \ h_{\bm R \bm R'}^{\alpha \beta}(\bm A) \ c_{\bm R',\beta}
\end{align}
where $c_{\bm R, \alpha}$ is the annihilation operator for an electron at the site $\bm R$ and orbital $\alpha$, and $\langle\bm R, \bm R'\rangle$ indicates a sum over nearest neighbors sites.
The models are all similar in that there are two competing hopping terms, an intra-unit cell hopping term whose magnitude is given by $|w|$ and an inter-unit cell hopping term whose magnitude is given by $|t|$. 
The different models are characterized by the different number of orbitals per unit cell, number of nearest neighbors, and hopping patterns encoded in the hopping matrices $h_{\bm R \bm R'}$. 


\subsection{\texorpdfstring{Fourfold symmetric model with $3e/4$ and $e/2$ corner chares}{}}

\begin{figure*}
  \centering
  \subfloat[]{\includegraphics[]{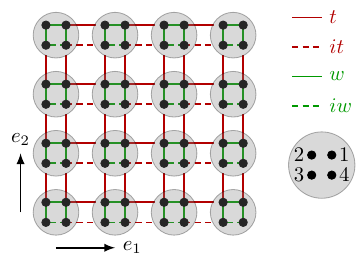}} 
  \hspace{15pt}
  \subfloat[]{\includegraphics[scale = 0.63]{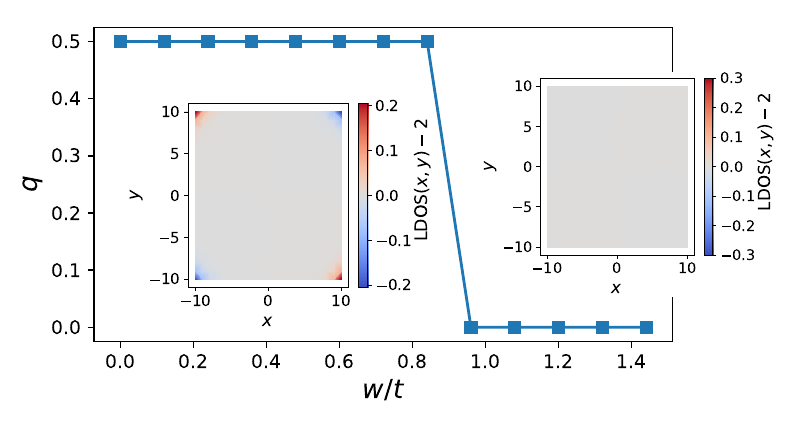}}

  \subfloat[]{\includegraphics[scale = 0.9, trim= -35 0 -35 0, clip]{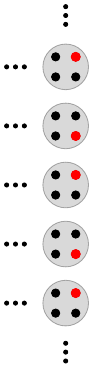}}
  \subfloat[]{\includegraphics[scale = 0.63]{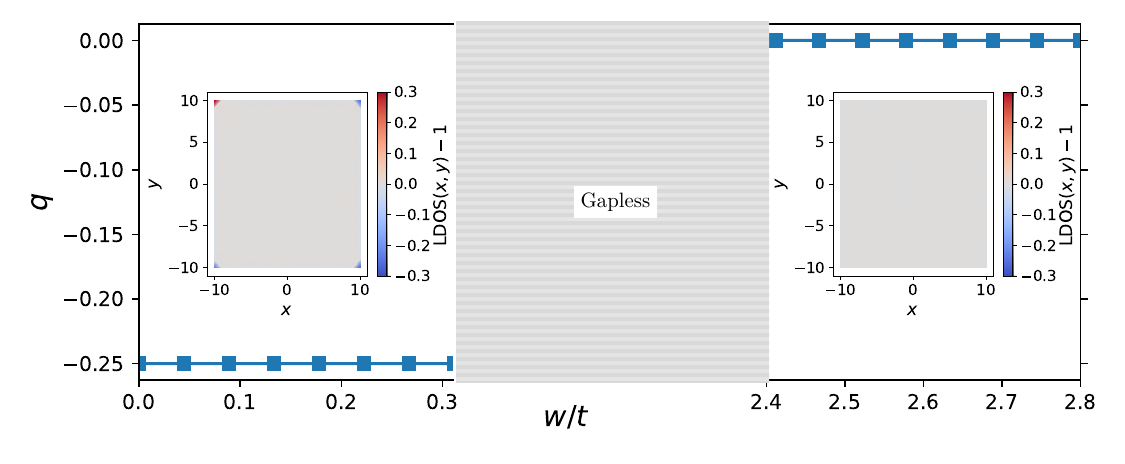}}

  \caption{Tight-binding model for a quadrupole insulator. In (a) the hopping terms are defined. We use a $\pi/2$ flux per plaquette  to study the model at both half and quarter-filling. Panel (b) shows the topological invariant evaluated for the model at half-filling, and the insets show the local density of states (LDOS) for the two phases of the model. The invariant correctly captures the phases and its value predicts the correct corner charges as in Eq.~(\ref{eq:17}). At quarter-filling with open boundary conditions, the model has gapless edge modes which we get rid of by applying a staggered potential on the edges in a fourfold rotation symmetric way as shown in (c). Orbitals on the edge marked by red has a positive on-site potential energy. Such a staggered potential is physically equivalent to inducing a charge density wave on the edge to make sure it is gapped. 
  In panel (d) we calculate the invariant at quarter-filling and also show the corner modes for the gapped phases. Again the value of the invariant correctly corresponds to the corner charges. 
  }
  \label{fig:bbh_model_def}
\end{figure*}

The prototypical example of an insulator hosting $e/2$ corner charges is the BBH model which is a four-band model defined on a square lattice.
Originally the model was defined with a $\pi$ flux per plaquette~\cite{Benalcazar_2017}. 
This leads to the bottom two bands being degenerate, and when filled the model has $e/2$ corner charges. 
We modify the model in a way that makes filling just one band of the model more natural, by threading a $\pi/2$ flux instead of $\pi$ per plaquette, putting all four bands of the model at different energies. 
A $C_4$-symmtric version is shown in Fig.~\ref{fig:bbh_model_def}(a), where every grey circle represents a unit cell and the black nodes represent the different orbitals. We take the positions of the orbitals to be at the centers of the unit cells, but draw them at different positions in Fig.~\ref{fig:bbh_model_def}(a) for better visual representation.  

The hopping matrices for the $\pi/2$ BBH model can be written explicitly as,
\begin{align}
    \quad h_{\bm R \bm R}(\bm A) &= 
    \begin{bmatrix}
        0 & 1 & 0 & 1 \\ 
        1 & 0 & 1 & 0 \\ 
        0 & 1 & 0 & -i\\ 
        1 & 0 & i & 0
    \end{bmatrix}, \nonumber \\ 
    h_{\bm R + e_1, \bm R}(\bm A) &= h^{\dagger}_{\bm R - e_1, \bm R}(\bm A)  \nonumber  \\  
    &  = \exp\[i\phi_{\bm R + e_1, \bm R}(\bm A)\]\begin{bmatrix}
        0 & 0 & 0 & 0 \\ 
        1 & 0 & 0 & 0 \\
        0 & 0 & 0 & -i \\ 
        0 & 0 & 0 & 0
    \end{bmatrix} , \label{eq:28}\\ 
    h_{\bm R + e_2, \bm R} (\bm A) &= h^\dagger_{\bm R - e_2, \bm R}(\bm A) \nonumber \\ 
    &= \exp\[i\phi_{\bm R + e_2, \bm R}(\bm A)\]\begin{bmatrix}
        0 & 0 & 0 & 0 \\
        0 & 0 & 0 & 0 \\
        0 & 1 & 0 & 0 \\ 
        1 & 0 & 0 & 0
    \end{bmatrix},\nonumber 
\end{align}
where the primitive lattice vectors $e_1$ and $e_2$, and the orbitals numbering are defined on the right in Fig.~\ref{fig:bbh_model_def}(a). 
The phases $\phi_{\bm R' \bm R}(\bm A)$ are calculated using the Peierls substitution in the case of bulk hopping terms and through the prescription described in Fig.~\ref{fig:def_of_oboundary_phase_hop} in the case of hopping terms crossing the boundary of the sample. 

The model here has a fourfold rotation symmetry $C_4: \bm R = a e_1 + b e_2 \rightarrow -b e_2 + a e_1$ with its action on the orbitals given by
\begin{align}
    c_4 = \begin{bmatrix}
        0 & 0 & 0 & 1 \\ 
        1 & 0 & 0 & 0 \\
        0 & 1 & 0 & 0 \\ 
        0 & 0 & i & 0
    \end{bmatrix}. 
\end{align}
We note that since we use the symmetric gauge to add the magnetic field this form of the symmetry is the same even for non-zero $\bm A$. 

At half-filling, the model has two phases. 
For $|w/t| < 1$ it is in the topological phase, and for $|w/t| > 1$ it is in the trivial phase.~\cite{Benalcazar_2017}
In Fig.~\ref{fig:bbh_model_def}(b), we numerically evaluate our invariant $q$ using periodic boundary condition as discussed in Sec.~\ref{sec:def_inv}. 
The results show that the invariant does capture the topological phase transition of the model by suddenly changing value at the phase transition. According to Eq.~\eqref{eq:17}, the value $q=1/2$ in the topological phase predicts that the system under open boundary condition exhibits fractional corner charge $e/2 \mod e$.

In the insets of Fig.~\ref{fig:bbh_model_def}(b), we  plot the local density of states (LDOS) using open boundary condition for a representative point in each phase of the model. Working in a basis where there is no entanglement between different corners, we pick one ground state and plot the LDOS.  The values of the LDOS confirm that the model indeed hosts $\pm e/2$ corner charges.

We also solve the model and its invariant at  quarter filling. The invariant along with the corner charges are calculated in Fig.~\ref{fig:bbh_model_def}(d).  From the results, we see that the model has three phases as we increase the ratio $|w/t|$. 
For $|w/t| \ll 1$ the model is in the topological phase and host corner charges. In particular, $q=-1/4$, which predicts corner charges $\nu=3e/4 \mod e$.
As $|w/t|$ is increased, the band widths increase, and the lower two bands overlap, leading to a gapless phase at quarter-filling. 
For $|w/t|\gg 1$ the band widths decrease once more, and we obtain a gapped phase that is trivial with no corner charges. 

We compute the fractional corner charges from LDOS under open boundary condition. One subtlety here is that at quarter filling the bulk has a non-zero polarization that leads to gapless edges. 
This can be resolved by adding a staggered potential on the edge unit cells while still preserving the $C_4$ symmetry in such a way as to open the edge gap and make it neutral. As shown in Fig.~\ref{fig:bbh_model_def}(c), the orbitals on the edge marked in red have an additional positive potential energy shift. Physically, this can be thought of as inducing an edge charge density wave (CDW) in a fourfold symmetric way that cancels the edge charges due to the bulk polarization. In the insets of Fig.~\ref{fig:sixfold_symm_models}(c) we plot the LDOS under open boundary condition in the topological and trivial phases.
The LDOS for a ground state confirms that the model has $3e/4$ corner charges in the topologiclal phase which is captured by the invariant as expected from Eq.~(\ref{eq:17}).

\subsection{\texorpdfstring{Sixfold symmetric models with $e/2$, $2e/3$, and $e/6$ corner charges}{}}
\begin{figure*}
  \centering
  \subfloat[]{\includegraphics[scale=0.8]{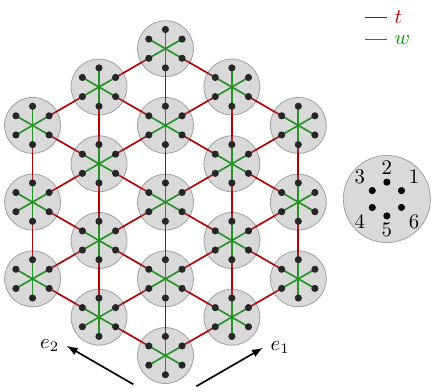}}
  \hspace{40pt}
  \subfloat[]{\includegraphics[scale=0.8]{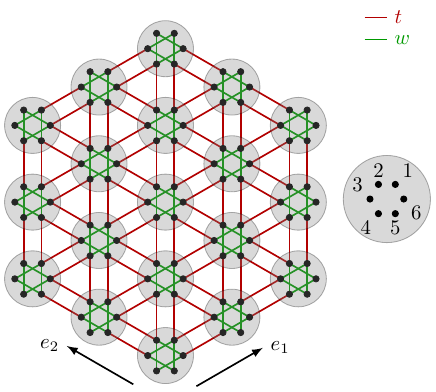}}

  \subfloat[]{\includegraphics[scale = 0.63]{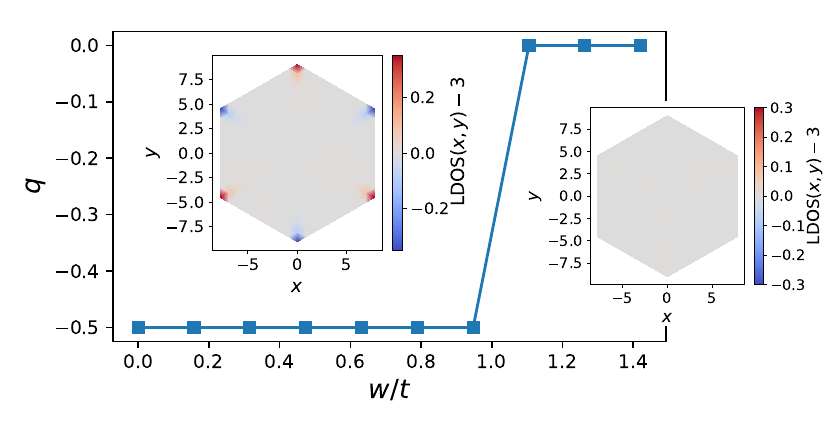}}
  \subfloat[]{\includegraphics[scale = 0.63]{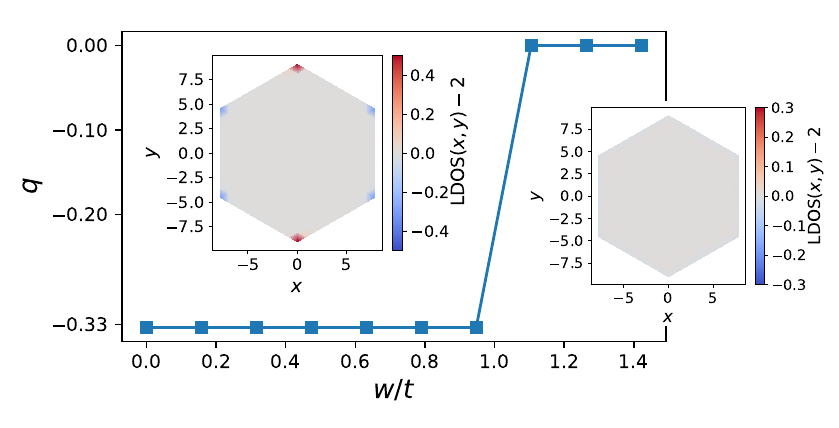}}
  \caption{We calculate our topological invariant for two different $C_6$ models that host corner charges. The model shown in (a) has a filling of three electrons per unit cell to ensure the bulk and edges are neutral. This model hosts $\pm e/2$ charges at the corners for $w/t > 1$. The model in (b) has a filling of two electrons per unit cell to achieve charge neutrality in the bulk and edges. This model hosts $-2e/3, \ e/3$ corner charges for $w/t > 1$.}
  \label{fig:sixfold_symm_models}
\end{figure*}

The sixfold symmetric models are defined on a triangular lattice. 
They have a similar structure to the BBH model with intra-unit-cell hopping terms that are proportional to $w$ and inter-unit-cell hopping terms that are proportional to $t$. 
Specific models hosting $e/2$ and $3e/2$ corner charges were proposed in Ref.~\cite{Benalcazar_2019}, and are defined in Figs.~\ref{fig:sixfold_symm_models}(a) and (b) respectively.


The sixfold rotation symmetry $C_6: \bm R = a e_1 + b e_2 \rightarrow (a-b)e_1 + a e_2$ with the orbitals transforming as the following, 
\begin{align}\label{eq:c6_on_orbitals}
    c_6 = \begin{bmatrix}
    0&0&0&0&0&1 \\
    1&0&0&0&0&0 \\
    0&1&0&0&0&0 \\ 
    0&0&1&0&0&0 \\
    0&0&0&1&0&0 \\
    0&0&0&0&1&0
    \end{bmatrix}
\end{align}

As discussed in Sec.~\ref{sec:def_inv}, we calculate the invariant on a hexagonal sample under periodic boundary condition implemented by identifying opposite sides of the hexagon. 
This treatment is essential for maintaining $C_6$ in the presence of non-zero vector potential. 
We consider the model in Fig.~\ref{fig:sixfold_symm_models}(a) at half filling, and the one in Fig.~\ref{fig:sixfold_symm_models}(b) at $1/3$ filling, both of which can be straightforwardly expressed in the form of tight-binding Hamiltonian in Eq.~\eqref{eq:27}.
The invariants and the resulting corner charges under open boundary condition are calculated for both models as shown in Figs.~\ref{fig:sixfold_symm_models}(c) and (d). 
Similar to the BBH model at half-filling, the models have two phases that are determined by the ratio between intra and inter-unit-cell hoppings $|w/t|$.
For $|w/t| < 1$ the models are in the topological phase with different corner charges. 
As shown in Figs.~\ref{fig:sixfold_symm_models}(b) and (c), the invariant captures the phase transition of both models and takes different values for the topological phases of the two models reflecting the different corner charges consistent with Eq.~(\ref{eq:17}).

Finally, by simply combining the tight-binding models in Fig.~\ref{fig:sixfold_symm_models}(a) at half filling and in Fig.~\ref{fig:sixfold_symm_models}(b) at $1/3$ filling, we see straightforwardly the topological invariant in the topological phase has $q=1/6$ and exhibits $e/6$ corner charge.

\section{Conclusion}\label{sec:conclusion}

In this work, we proposed a  $\mathbb{Z}_n$ many-body topological invariant for 2d second-order topological insulators with $C_n$ rotation symmetry. Our construction does not rely on multipolar moments in crystalline systems, which we argued to be inadequate to capture the second-order topology for a general $n$. Via a UV-IR matching we directly demonstrated the correspondence between the topological invariant in the bulk and the fractional electric charges at the corners quantized to be multiples of $e/n$. We also clarified the connection between our invariant with the generalized Wen-Zee shift for $C_n$-symmetric systems.

Compared to the various topological invariants proposed for higher-order topological insulators, our result readily applies in the presence of interactions, for a generic $C_n$ rotation symmetry, and even in the absence of translation symmetry. In particular, it will be interesting to compute the topological invariant in this work for quasicrystal systems, some of which has been recently shown to exhibit fractional corner charges.

It will be interesting to extend the topological invariant here to other higher-order topological phases, such as higher-order topological superconductors or higher-dimensional systems. For a higher-order topological superconductor, the magnetic flux are trapped in vortices, and it has been shown recently for noninteracting systems there exists a correspondence~\cite{zhang2022} between corner Majorana modes and the Majorana bound states in the vortex. An extension of our topological invariant for superconductors may generalize the vortex-corner correspondence beyond noninteracting limit~\cite{you2022}. We leave these interesting issues to future work.

\acknowledgments
We thank J. Herzog-Arbeitman, M. Barkeshli, G. Y. Cho, T. L. Hughes, M. Oshikawa, Y. Tada, Y. You,  and R.-X. Zhang for useful discussions. AJ and YW are supported by NSF under award number DMR-2045781. YML is supported by NSF under award number DMR-2011876. YW and YML acknowledges support by grant NSF PHY-1748958 to the Kavli Institute for Theoretical Physics (KITP), and NSF PHY-2210452 to the Aspen Center for Physics.
\bibliography{charge4e}

\end{document}